\documentclass[aps,prl,showpacs,amsmath,amssymb,amsfonts,lengthcheck,twocolumn,longbibliography]{revtex4-1}
\usepackage{graphicx}
\usepackage{subfigure}
\usepackage{amsthm}
\usepackage{verbatim}
\usepackage{dcolumn}
\usepackage{bm}
\usepackage{epsf}
\usepackage{color}
\usepackage[colorlinks=true,citecolor=blue,linkcolor=blue,urlcolor=blue]{hyperref}%
\usepackage{xcolor}
\usepackage{dsfont}
\newcommand{\bra}[1]{\left\langle #1\right|}
\newcommand{\ket}[1]{\left|#1\right\rangle}

\newcommand{\tr}[1]{\mathrm{tr}\left\{#1\right\}}

\newcommand{\la}{\left\langle}
\newcommand{\ra}{\right\rangle}

\newcommand{\td}{\mathrm{d}}

\newcommand{\e}[1]{\exp{\left(#1\right)}}
\newcommand{\lo}[1]{\ln{\left(#1\right)}}

\newcommand{\com}[2]{\left[#1,\,#2\right]}

\newcommand{\bla}{bla\\bla\\bla\\bla\\bla}

\newcommand{\PRA}{Phys. Rev. A }

\newcommand{\PRL}{Phys. Rev. Lett. }

\newcommand{\mc}[1]{\mathcal{#1}}

\newcommand{\mrm}[1]{\mathrm{#1}}

\DeclareMathOperator*{\sumint}{%
\mathchoice%
  {\ooalign{$\displaystyle\sum$\cr\hidewidth$\displaystyle\int$\hidewidth\cr}}
  {\ooalign{\raisebox{.14\height}{\scalebox{.7}{$\textstyle\sum$}}\cr\hidewidth$\textstyle\int$\hidewidth\cr}}
  {\ooalign{\raisebox{.2\height}{\scalebox{.6}{$\scriptstyle\sum$}}\cr$\scriptstyle\int$\cr}}
  {\ooalign{\raisebox{.2\height}{\scalebox{.6}{$\scriptstyle\sum$}}\cr$\scriptstyle\int$\cr}}
}

\begin{document}

\title{Quantum speed limits and the maximal rate of information production}

\author{Sebastian Deffner}
\email{deffner@umbc.edu}
\affiliation{Department of Physics, University of Maryland, Baltimore County, Baltimore, MD 21250, USA}

\date{\today}

\begin{abstract}
The Bremermann-Bekenstein bound sets a fundamental upper limit on the rate with which information can be processed. However, the original treatment heavily relies on cosmological properties and plausibility arguments. In the present analysis, we derive equivalent statements by relying on only two fundamental results in quantum information theory and quantum dynamics -- Fannes inequality and the quantum speed limit. As main results, we obtain Bremermann-Bekenstein-type bounds for the rate of change of the von Neumann entropy in quantum systems undergoing open system dynamics, and for the rate of change of the Shannon information encoded in some logical states undergoing unitary quantum evolution.
\end{abstract}

\maketitle

\subparagraph{Introduction}

All around the globe government funded research institutions, big corporations, as well as dedicated start-ups are pursuing the next technological breakthrough -- to achieve quantum supremacy \cite{Sanders2017,Riedel2017,Raymer2019}. In essence, quantum supremacy means that quantum technologies can perform certain tasks while consuming less resources than their classical analogs. Typically, such quantum advantage is expected to be found in quantum sensing, quantum communication, and quantum computing \cite{Raymer2019}. 

From the point of view of quantum thermodynamics \cite{Deffner2019book}, all three of these applications with possible quantum advantages have in common that one way or another entropy is produced and information is transferred. Therefore, the natural question arises whether the intricacies of quantum physics pose any fundamental limits on the rate of change of the entropy stored in quantum systems. Remarkably, this rather fundamental problem has been subject of investigation for more than five decades. 

In particular, Bremermann \cite{Bremermann1967} proposed that any computational device must obey the fundamental laws of physics namely \emph{special relativity}, \emph{quantum mechanics}, and \emph{thermodynamics}. Then, identifying Shannon's noise energy with the $\Delta E$ in Heisenberg's uncertainty relation \cite{Heisenberg1927} for energy and time, $\Delta E \Delta t\geq \hbar$, he found an upper bound on the rate with which information can be communicated. However, the rather cavalier identification of quantum uncertainty with the channel capacity in information theory could not and cannot be considered sound. 

Consequently, Bekenstein proposed an alternative approach \cite{Bekenstein1981} that was motivated by understanding the upper bound on the rate with which information can be retrieved from black holes \cite{Bekenstein1973,Bekenstein1974,Hawking1975,Bekenstein1981PRD,Bekenstein1990}. However, despite its rather appealing simplicity, Bekenstein's bound can also not be considered satisfactory as it is not clear whether or how the bound applies to problems of quantum communication. Also more conceptually, one would like to be able to estimate the rate of information transfer in a quantum computer without having to refer to black holes \cite{Bekenstein1973,Bekenstein1974,Hawking1975,Bekenstein1981PRD,Bekenstein1990}.

Interestingly, bounding the rate with which entropy and information can be communicated is a rather involved problem, that has been under constant attention since its first inception \cite{Pendry1983,Landauer1987,Bekenstein1988,Caves1994,Blencowe2000,Lloyd2004,Garbaczewski2007,Pei_Rong2010,Guo2011,Guo2012,Bousso2017,Lewis2019}. A particularly interesting application may be found in quantum thermodynamics, where a bound on the rate of entropy production would necessarily cause quantum devices to operate closer to equilibrium than classical devices  \cite{Deffner2010PRL}.

From Bremermann's original treatment it appears obvious that an origin of upper bounds on rates may be sought by considering the quantum speed limit \cite{Deffner2017JPA}. The quantum speed limit sets the minimal time a quantum system needs to evolve between distinguishable states, and as such is a rigorous treatment of Heisenberg's uncertainty relation for energy and time \cite{Heisenberg1927}. For a comprehensive treatment and its history we refer to a recent Topical Review \cite{Deffner2017JPA}. More recently, the quantum speed limit has found applications in a wide range of problems, including but not limited to metrology \cite{Giovannetti2011,Campbell2018}, quantum control \cite{Cimmarusti2015,Campbell2017,Funo2017}, thermodynamics \cite{Safranek2018,Funo2019}, and in studying the quantum to classical transition \cite{Shanahan2018,Shiraishhi2018}.

Thus, the natural question arises whether a sound and practical bound on the rate of quantum communication can be obtained by considering the quantum speed limit for entropy changes. In the present work, we will show that the Bremermann-Bekenstein bound is an immediate consequence of the Fannes inequality \cite{Fannes1973} and a judiciously chosen version of the quantum speed limit. Thus, we will show that the maximal rate of \emph{quantum information production} follows simply from combining two of the most fundamental statements in quantum information theory. More specifically, we will derive upper bounds on the rate of information production in finite- and infinite-dimensional Hilbert spaces, that is for microcanonical and canonical scenarios. Moreover, we will also bound the rate of information production over logical subspaces for unitary dynamics. 

\subparagraph{Preliminaries}

To establish notions and notations, we begin by briefly reviewing some seminal results from the literature.

\paragraph{Bremermann-Bekenstein bound} 

Critiquing Bremermann's original account \cite{Bremermann1967}, Bekenstein proposed an alternative derivation of an upper bound on the rate of entropy production \cite{Bekenstein1981}. His treatment starts by bounding the total amount of entropy, $S$, that can be stored in any given region of space, which can be expressed as \cite{Bekenstein1973,Hawking1975,Bekenstein1981PRD}
\begin{equation}
\label{eq:entropy_BH}
\frac{S}{E_0} \leq \frac{2\pi k_B R}{\hbar c}\,,
\end{equation}
where $E_0$ and $R$ are the internal energy and the effective radius of the system in its rest frame, respectively. Now noting that information, $\mc{I}$, and entropy are related by $\mc{I}=S/k_B \lo{2}$, and that no message can travel faster than the speed of light $c$, we have \cite{Bekenstein1981}
\begin{equation}
\mc{I}\leq \frac{2 \pi}{\gamma \hbar c \lo{2}}\,E R
\end{equation}
where $\gamma$ is the Lorentz factor corresponding to the relative motion of sender and receiver. Moreover, $E$ is now the energy of the message in the receiver's frame. 

Finally, Bekenstein noted that the minimal time for a message to cross the distance from sender to receiver is $\tau\geq 2 R/\gamma c$, and therefore the maximal rate of information transfer becomes
\begin{equation}
\label{eq:Bekenstein}
\dot{\mc{I}} \leq \frac{\pi E}{\hbar \lo{2}}\,,
\end{equation}
where we introduced the notation $\dot{\mc{I}}=\mc{I}/\tau$.

It is interesting to note, that Eq.~\eqref{eq:Bekenstein} can be interpreted as a dynamical version of Landauer's principle \cite{Landauer1961,Landauer1987} applied to information transfer. Remarkably the Hawking-Bekenstein entropy \eqref{eq:entropy_BH} also led to development of black hole thermodynamics and, eventually, the formulation of the information paradox. Since this issue seems still hotly debated in cosmology \cite{Mathur2009}, Eq.~\eqref{eq:Bekenstein} might not to be taken too literally. For instance, it is still nor entirely clear how to determine the energy $E$ of a black hole that determines the rate with which information is emitted \cite{Johnson2014}. Moreover, it is not immediately clear whether and how the Bremermann-Bekenstein bound applies to communicating quantum information. 

\paragraph{Pendry's bound}

Addressing some of these issues, Pendry proposed yet another solution to the problem \cite{Pendry1983}. Pendry noted that the time for a receiver to detect $N$ bits of information, quantum mechanics dictates an uncertainty of energy $\delta E$ to be no less than, $\delta E\sim \hbar/t$. If information is now conveyed by whether or not a particle of a given energy has arrived, we have
\begin{equation}
\dot{E}\geq \frac{1}{2}\frac{\hbar N^2}{t^2}\,.
\end{equation}
The latter inequality assumes that each particle carries an average energy of $N\hbar/2 t$, such that the total energy spread of the message is $N\delta E$.

A more precise analysis, involving an average of the Bose-Einstein distribution of the information carriers, it can be shown \cite{Pendry1983} that
\begin{equation}
\label{eq:Pendry}
\dot{\mc{I}}^2\leq \frac{\pi}{3\hbar\,\left(\lo{2}\right)^2}\,\dot{E}\,,
\end{equation} 
where $\dot{\mc{I}}=N/t$.

The obvious advantage of Pendry's bound \eqref{eq:Pendry} is that its derivation entirely relies on arguments from quantum statistical mechanics. Remarkably, the only essential ingredient is Heisenberg's uncertainty relation for energy and time. Due to the recent progress in understanding the quantum speed limit \cite{Deffner2017JPA}, it appears thus only natural to revisit the maximal rate of quantum information transfer.

\paragraph{Quantum speed limit from trace distance}

The \emph{quantum speed limit time} is the characteristic, minimal time a quantum system need to evolve between distinguishable states. Generally this evolution is described by a Master equation, $\dot{\rho}=L(\rho)$, which can be unitary or dissipative, driven or undriven, linear or nonlinear. Typically, the distinguishability of quantum states is measured by some measure on density operator space such as, for instance, the Bures angle \cite{Deffner2013PRL,Taddei2013}, the relative purity \cite{Campo2013}, or the Wigner-Yanase skew information \cite{Pires2016,Brody2019}. 

For the present purposes, and for the sake of simplicity, we will be working with the trace distance between initial state $\rho_0$ and time evolved state $\rho_t$ \cite{Deffner2017NJP},
\begin{equation}
\label{eq:trace}
\ell(\rho_t,\rho_0)=\frac{1}{2}\tr{\left|\rho_t-\rho_0\right|}\,.
\end{equation}
The rate of change $\dot{\ell}$ can then be upper bounded, and we have (under slight abuse of notation)
\begin{equation}
\dot{\ell}\leq|\dot{\ell}|=\frac{1}{2}\left|\tr{\frac{\left(\dot{\rho}_t(\rho_t-\rho_0)+(\rho_t-\rho_0)\dot{\rho}_t\right)}{2\left|\rho_t-\rho_0\right|}\,}\right|\,.
\end{equation}
Further employing the triangle inequality, we immediately obtain
\begin{equation}
\dot{\ell}\leq \frac{1}{2}\,\tr{\left|\dot{\rho}_t\right|}\equiv \frac{1}{2}\,||\dot{\rho}_t||\,.
\end{equation}
Now using the standard arguments \cite{Deffner2013PRL,Deffner2017JPA,Deffner2017NJP} and integrating over an interval of length $\tau$, we have
\begin{equation}
\ell(\rho_\tau,\rho_0)\leq \frac{\tau}{2}\,\Lambda_\tau\,,
\end{equation}
where $\Lambda_\tau$ is the time averaged trace norm, $\Lambda_\tau=1/\tau\,\int_0^\tau dt\, ||\dot{\rho}||$ \cite{Deffner2013PRL,Deffner2017NJP}. Thus, the quantum speed limit time can be written as
\begin{equation}
\label{eq:QSL}
\tau_\mrm{QSL}=\frac{2\,\ell(\rho_\tau,\rho_0)}{\Lambda_\tau}\,.
\end{equation}

It is important to note that Eq.~\eqref{eq:QSL} is \emph{not} the sharpest bound on the minimal quantum evolution time. However, the following, mathematical analysis becomes particularly simple for the quantum speed limit time \eqref{eq:QSL} based on the trace distance \eqref{eq:trace}. 

Before we continue it is also interesting to note that Eq.~\eqref{eq:QSL} reduces to the standard Heisenberg uncertainty relation for energy and time under the appropriate assumptions \cite{Deffner2013PRL,Deffner2017JPA}. In particular, we have for unitary dynamics governed by positive semi-definite, constant Hamiltonians, $\dot{\rho}=1/i\hbar\, \com{H}{\rho_t}$, and for special initial states with $\com{H}{\rho_0}=0$,
\begin{equation}
\tau_\mrm{QSL}=\frac{\hbar}{E}\,\ell(\rho_\tau,\rho_0)\,.
\end{equation}
The latter is nothing else but the Margolus-Levitin bound \cite{Margolus1998}, where we measure angles in units of radian/$(\pi/2)$.

In a more general case, the physical interpretation of  $\Lambda_\tau$ is less immediate. In the literature, it has proven successful the consider $\Lambda_\tau$ as the average quantum speed \cite{Deffner2017JPA} that characterizes all dynamical properties of a quantum system \cite{Campbell2017,Campbell2018,Fogarty2019} \footnote{With the help of the triangle inequality it becomes apparent that $\Lambda_\tau$ is upper bounded by the time-averaged energy and by an contribution from the interaction with the environment.}.

\subparagraph{Non-unitary dynamics}

The informational content of an arbitrary quantum state, $\rho$, is given by its von Neuman entropy \cite{Nielsen2010}
\begin{equation}
\label{eq:entropy}
\mc{H}(\rho)=-\tr{\rho\lo{\rho}}\,.
\end{equation}
For the present purposes, we then consider the processing and/or communicating of quantum information to be determined by a change of $\mc{H}$. Thus, we are interested in bounding the rate of change $\dot{\mc{H}}$, as the quantum system undergoes arbitrary, non-unitary dynamics described by some master equation, $\dot{\rho}=L(\rho)$. Note that,  as before, we do not pose any restrictions on the quantum Liouvillian $L$, and hence the dynamics is explicitly allowed to be Markovian or non-Markovian, and also linear as well as nonlinear.

\paragraph{The microcanonical bound}

For the sake of simplicity we start with the situation in which $\rho$ lives in finite-dimensional Hilbert spaces. For such quantum states, the von Neumann entropy  \eqref{eq:entropy} fulfills the seminal Fannes inequality \cite{Fannes1973,Audenaert2007,Nielsen2010}
\begin{equation}
\label{eq:Fannes}
\left|\mc{H}(\rho)-\mc{H}(\sigma)\right|\leq 2\,\ell (\rho,\sigma) \lo{d}+1/e\,,
\end{equation}
where $\ell(\rho,\sigma)=1/2\,\tr{\left|\rho-\sigma\right|}$ is again the trace distance. It is worth emphasizing that Eq.~\eqref{eq:Fannes} can be generalized to asymptotically tight bounds \cite{Audenaert2007}. However, for the present treatment we will be working with the simplest mathematical representation to keep the physical interpretation as transparent as possible. Tightening the bounds will then be a simple exercise.

Now, consider two quantum states $\sigma$ and $\rho$, such that $\rho_0=\rho$ and $\rho_\tau=\sigma$ under the dynamics described by $\dot{\rho}=L(\rho)$. In this case  Eq.~\eqref{eq:Fannes} simply becomes
\begin{equation}
\label{eq:inequality_S}
\left|\Delta \mc{H}\right|\leq \lo{d}\,\tau_\mrm{QSL}\,\Lambda_\tau+1/e\,,
\end{equation}
where we replaced the trace distance, $\ell(\rho_\tau,\rho_0)$, with the quantum speed limit time \eqref{eq:QSL}. Note that for large enough dimensions, $d\gg1$ the second term in Eq.~\eqref{eq:inequality_S} becomes negligible. Then, further defining $\dot{\mc{I}}\equiv \left|\Delta \mc{H}\right|/\tau_\mrm{QSL} \lo{2}$ we can write
\begin{equation}
\label{eq:bound_1}
\dot{\mc{I}}\lesssim \frac{\lo{d}}{\lo{2}}\,\Lambda_\tau\,,
\end{equation}
which is a fundamental upper bound on the rate of change of the von Neumann entropy. 

It is worth stressing that Eq.~\eqref{eq:bound_1} sets an upper bound on the rate of quantum information production, which is related by not identical to entropy production in quantum thermodynamics \cite{Deffner2019book}. In a thermodynamic context, entropy production would be defined as the difference between the change of von-Neumann entropy and the heat exchanged with the environment, $\Sigma=\Delta H-\beta Q$, where $\beta$ is the inverse temperature.

Equation~\eqref{eq:bound_1} can be interpreted as a version of Eq.~\eqref{eq:Bekenstein}. As we have seen above, $\Lambda_\tau$ generalizes the energy $E/\hbar$ as the determining factor of the maximal speed in open system dynamics. Moreover, the logarithm of the dimension of the accessible Hilbert space, $\lo{d}$, is often identified as the microcanonical, Boltzmann entropy \cite{Deffner2016NJP}, $\mc{S}_B=k_B \lo{d}$. Therefore, we can also write
\begin{equation}
\label{eq:bound_2}
\dot{\mc{I}}\lesssim \frac{\mc{S}_B\,\Lambda_\tau}{k_B\,\lo{2}}\,,
\end{equation}
which is the quantum information theoretic version of the Bremermann-Bekenstein bound~\eqref{eq:Bekenstein}. 

We emphasize that Eq.~\eqref{eq:bound_2} was obtained rigorously by combining only two fundamental results in (i) quantum information theory, the Fannes inequality \eqref{eq:Fannes}, and (ii) quantum dynamics, the quantum speed limit \eqref{eq:QSL}. No plausibility arguments, and also no reference to the properties of black holes was necessary.

Note, that the present analysis is phrased in its mathematically simplest and most appealing form. Thus, Eq.~\eqref{eq:bound_2} is not the tightest possible bound on the rate of quantum communication. Tighter bounds can be obtained with the help of sharpened Fannes-type inequalities \cite{Audenaert2007}, and Riemannian formulations of the quantum speed limit \cite{Pires2016,Deffner2017NJP,Deffner2017JPA}.

\paragraph*{Example: damped Jaynes-Cummings model}

\begin{figure}
\includegraphics[width=.48\textwidth]{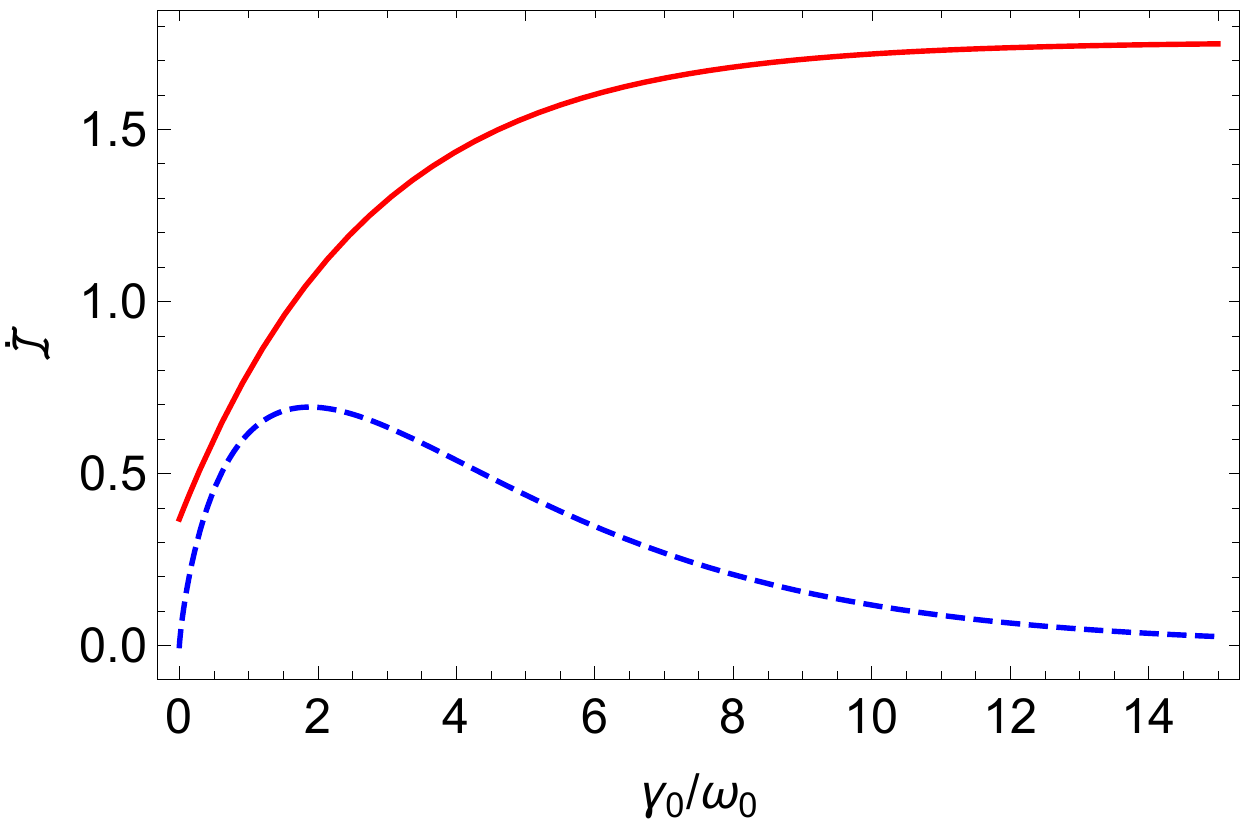}
\caption{\label{fig:micro} Exact, average rate of information production (blue, dashed line), and upper bound as determined by Eq.~\eqref{eq:bound_2}. Note that additive term in Eq.~\eqref{eq:inequality_S}, $1/e$, is not small enough to be neglected in this case. Parameters are $\lambda=1$, $d=1$, and $\tau=1$.}
\end{figure}

Before we continue we illustrate and verify Eq.~\eqref{eq:bound_2} with a pedagogical example. To this end, we analyze a two-level atom dissipatively coupled to a leaky, optical cavity. This model is commonly known as damped Jaynes-Cummings model \cite{Breuer2007}, and it is fully analytically solvable. The exact master equation for the reduced density operator of the qubit, $\rho_t$, can be written as \cite{Breuer2007,Garraway1997}
\begin{equation}
\begin{split}
\dot{\rho}_t=&-\frac{i}{\hbar} \com{H_\mrm{qubit}}{\rho_t}-\frac{i}{2\hbar} \com{\lambda_t\, \sigma_+\,\sigma_-}{\rho_t}\\
+&\gamma_t\left(\sigma_-\rho_t\sigma_+ -\frac{1}{2}\sigma_+\sigma_-\rho_t-\frac{1}{2}\rho_t\sigma_+\sigma_-\right)\,,
\end{split}
\end{equation}
where $H_\mrm{qubit}=\hbar \omega_0\,\sigma_+\sigma_-$. The time-dependent decay rate, $\gamma_t$, and the time-dependent Lamb shift, $\lambda_t$, are fully determined by the spectral density, $J(\omega)$, of the cavity mode. We have
\begin{equation}
\label{q13}
\lambda_t=-2\,\mrm{Im}\left\{\frac{\dot{c}_t}{c_t}\right\} \quad \mathrm{and} \quad \gamma_t=-2\,\mrm{Re}\left\{\frac{\dot{c}_t}{c_t}\right\}\,
\end{equation}
where $c_t$ is a solution of
\begin{equation}
\label{q14}
\dot{c}_t=-\int_0^t\td s\int \td \omega\,J(\omega)\,\e{i\hbar\left(\omega-\omega_0\right)(t-s)}\,c_s\,.
\end{equation}
This model has been extensively studied, since it is exact and completely analytically solvable \cite{Breuer2007,Garraway1997}. Moreover, it is of thermodynamic relevance as it allows the study of non-Markovian quantum dynamics \cite{Breuer2009,Laine2010,Xu2010,Hou2011,Fonseca2012,Deffner2013PRL,Deffner2014} and it has been realized in a solid-state cavity QED \cite{Madsen2011}.

Further assuming that there is only one excitation in the combined atom-cavity system, the environment can be described by an effective Lorentzian spectral density of the form,
\begin{equation}
J(\omega)=\frac{1}{2\pi}\frac{\gamma_0\lambda}{(\omega_0-\omega)^2+\lambda^2},
\end{equation}
where $\omega_0$ denotes the frequency of the two-level system, $\lambda$ the spectral width and $\gamma_0$ the coupling strength. The time-dependent decay rate is then explicitly given by,
\begin{equation}
\gamma_t = \frac{2\gamma_0 \lambda \sinh(dt/2)}{d\cosh(dt/2) + \lambda \sinh(dt/2)},
\end{equation}
where $d=\sqrt{\lambda^2-2\gamma_0\lambda}$.

It is then a simply exercise to compute the time averaged rate of change of the von Neumann entropy, and the upper bound as given in Eq.~\eqref{eq:bound_1} explicitly. For a qubit that is initially in its ground state, $\rho_0=\ket{0}\bra{0}$, the results are depicted in Fig.~\ref{fig:micro}. We observe that in the highly non-Markovian limit, $\gamma_0/\omega_0 \gg 1$ the upper bound is finite, whereas the actual, average rate of information production vanishes. This is to be expected, as for highly non-Markovian dynamics the von Neumann entropy strongly oscillates, whereas the master equation itself depends only weakly on time \cite{Breuer2009}. However, we also observe that in the Markovian limit the upper bound is reasonably close to the exact value of the information production despite the fact that the employed inequalities are not tight.

\paragraph{The canonical bound}

An obvious shortcoming of above treatment is that the Fannes inequality \eqref{eq:Fannes} is only meaningful for finite-dimensional Hilbert spaces. This was remedied by Winter \cite{Winter2016}, who generalized Eq.~\eqref{eq:Fannes} to infinite dimensional quantum systems with bounded energy. 

To this end, consider a system with finite average energy, $\la H\ra \leq E$. Then the (unique) state that maximizes the von-Neumann entropy is given by the Gibbs state, $\rho_\mrm{eq}(E)=\e{-\beta H}/Z$ with  $\tr{\e{-\beta H}\left(H-E\right)}=0$, and where $Z=\tr{\e{-\beta H}}$ is the canonical partition function. In this case the Fannes inequality \eqref{eq:Fannes} can be generalized to read \cite{Winter2016}
\begin{equation}
\label{eq:Fannes_can}
\left|\mc{H}(\rho)-\mc{H}(\sigma)\right|\leq 2 \ell(\rho,\sigma)\,\mc{H}(\rho_\mrm{eq}(E/\ell))+2\lo{2}\,,
\end{equation}
where we replaced the second term from Ref.~\cite{Winter2016} with its maximal value.

Introducing the Gibbs entropy $\mc{S}_G=\beta (E-F)$, where as always $F=-1/\beta \lo{Z}$, we immediately obtain 
\begin{equation}
\label{eq:bound_3}
\dot{\mc{I}}\lesssim \frac{\mc{S}_G\,\Lambda_\tau}{k_B\,\lo{2}}\,,
\end{equation}
and, as above $\dot{\mc{I}}\equiv \left|\Delta \mc{H}\right|/\tau_\mrm{QSL} \lo{2}$. Note that as before we suppressed the small additive term, which is a fair approximation for, here, large enough energies $E$.

Equations~\eqref{eq:bound_2} and \eqref{eq:bound_3} constitute our main results for general, non-unitary quantum dynamics. For finite dimensional as well as for infinite dimensional Hilbert spaces, the rate of information production can be bounded with the help of the quantum speed limit \eqref{eq:QSL}. However, many problems in quantum computing are designed from unitary quantum dynamics \cite{Nielsen2010}. Thus Eq.~\eqref{eq:bound_2} and \eqref{eq:bound_3} are not very instructive for practical purposes.

\subparagraph{Unitary dynamics -- Shannon information}

Fortunately, the above framework can be easily generalized to information dynamics described by unitary quantum evolution. For isolated quantum systems, one is often interested in the Shannon information in the eigenbasis of a particular observable, $\mc{X}$,
\begin{equation}
S_\mc{X}(\rho)=-\sumint \rho(x)\lo{\rho(x)}\,,
\end{equation}
where $\rho(x)=\bra{x}\rho\ket{x}$ is the marginal distribution over $\mc{X}$, and $\ket{x}$ are the eigenstates of $\mc{X}$. If $\mc{X}$ is the energy of the system, $E$, the Shannon information $S_E$ is called the diagonal entropy, which has been shown to be of thermodynamic significance \cite{Polkovnikov2011}. However, also more generally if only partial information is accessible, the dynamics of $S_\mc{X}$ can be of fundamental importance \cite{Garbaczewski2007}. As an example, consider time-of-flight experiments with Bose-Einstein condensates \cite{Ketterle2002}, in which only the position distribution can be measured, but full quantum state tomography is not available. Finally, in a quantum computational setting $\ket{x}$ can be thought of as ``logical states'', which are a quantum version of ``information bearing degrees of freedom'', cf. Ref.~\cite{Deffner2013PRX} for a classical treatment.

It has been shown rather recently \cite{Polyanskiy2016}  that the Shannon information fulfills a continuity bound that strongly resembles Fannes inequality \eqref{eq:Fannes},
\begin{equation}
\left|S_\mc{X}(\rho)-S_\mc{X}(\sigma)\right|\leq \alpha\, \mc{W}_2(\rho(x),\sigma(x))
\end{equation}
where $\mc{W}_2(\rho,\sigma)$ is the Wasserstein-2 distance \cite{Wasserstein1969}. Generally, the Wasserstein-$p$ distance reads
\begin{equation}
\mc{W}_p(\rho,\sigma)=\left(\sumint\left|\bra{x}\rho\ket{x}-\bra{x}\sigma\ket{x}\right|^p\right)^{1/p}\,,
\end{equation}
and thus $\mc{W}_p(\rho,\sigma)$ is equivalent to the Schatten-$p$ distance extended to continuous probability space. Furthermore, the number $\alpha$ is entirely determined by the second moments of the distributions $\rho(x)$ and $\sigma(x)$ \cite{Polyanskiy2016}
\begin{equation}
\alpha=c_1 \left(\sqrt{\la x^2\ra_\rho}+\sqrt{\la x^2\ra_\sigma}\right)+c_2\,,
\end{equation}
where $c_1>0$ and $c_2>$ are constants dependent on the choice of $\mc{X}$. Note that the Wasserstein-$p$ distances fulfill the same ordering as the Schatten-$p$ distances, and in particular we have 
\begin{equation}
\mc{W}_2(\rho,\sigma)\leq \mc{W}_1(\rho,\sigma)\,.
\end{equation}
Thus we can also write,
\begin{equation}
\left|S_\mc{X}(\rho)-S_\mc{X}(\sigma)\right|\leq \alpha\, \mc{W}_1(\rho(x),\sigma(x))\,,
\end{equation}
which is mathematically more appealing for the remaining analysis.

\paragraph{Quantum speed limit for marginals}

We have recently shown that quantum speed limits can also be determined from the rate of change of the Wigner function \cite{Deffner2017NJP}. Analogous arguments apply to the dynamics of the marginals. To this end, consider now a quantum system that evolves under unitary, von Neumann dynamics, $\dot{\rho}=1/i\hbar\,\com{H}{\rho}$. Then, $\mc{W}_1(\rho_t(x),\rho_0(x))$ measures how far the distribution of observable values of $\mc{X}$ travels from their initial values. 

In particular, we also have
\begin{equation}
\dot{\mc{W}}_1\leq \left|\dot{\mc{W}}_1\right|=\left|\sumint \frac{\rho_t(x)-\rho_0(x)}{\left|\rho_t(x)-\rho_0(x)\right|}\,\dot{\rho}_t(x)\right|\,,
\end{equation}
which reduces with the help of the triangle inequality to
\begin{equation}
\dot{\mc{W}}_1\leq \sumint \left|\dot{\rho}_t(x)\right|=||\dot{\rho}_t(x)||_1\,.
\end{equation}
The latter inequality can be used to define a quantum speed limit time in the space spanned by the eigenstates of $\mc{X}$. We obtain
\begin{equation}
\label{eq:QSL_W}
\tau_\mrm{QSL}^\mc{X}\equiv\frac{\mc{W}_1(\rho_\tau(x),\rho_0(x))}{\Lambda^\mc{X}_\tau}\,,
\end{equation}
where as before $\Lambda^\mc{X}_\tau=1/\tau \,\int dt\, ||\dot{\rho}_t(x)||_1$

\paragraph{Bekenstein-type bound for Shannon information}

The quantum speed limit time for marginal distributions \eqref{eq:QSL_W} then allows to derive a Bekenstein-type bound for the change of Shannon information. In complete analogy to before, we define $\dot{\mc{I}}_\mc{X}\equiv |\Delta S_\mc{X}|/\tau_\mrm{QSL}^\mc{X} \lo{2}$, and thus we have
\begin{equation}
\label{eq:bound_4}
\dot{\mc{I}}_\mc{X}\leq \frac{\alpha\, \Lambda^\mc{W}_\tau}{\lo{2}}\,.
\end{equation}

Comparing the original Bremermann-Bekenstein bound \eqref{eq:Bekenstein}, our bounds for non-unitary dynamics \eqref{eq:bound_2} and \eqref{eq:bound_3}, and the last bound for the rate of change of the Shannon information in unitary dynamics \eqref{eq:bound_4}, we observe a mathematically universal form. For any kind of information processing including communication, the maximal rate is determined by the quantum speed limit \cite{Deffner2017JPA} and a situation dependent prefactor.

\subparagraph{Concluding Remarks}

In the present analysis we have obtained several upper bounds on the rate of quantum information production, for quantum systems undergoing (i) nonunitary dynamics in finite Hilbert spaces, (ii) nonunitary dynamics in infinite dimensional Hilbert spaces, and (iii) for the information carried by marginal distributions under unitary dynamics. Remarkably, in all of these situations we recovered the same mathematical form of the upper bound as proposed by Bekenstein by plausibility arguments involving the thermodynamic properties of black holes. However, our analysis is mathematically rigorous and relies on only two fundamental results in quantum information and dynamics, the continuity of quantum entropy and the quantum speed limit. Thus, we anticipate applications in virtually all areas of quantum physics, including but not limited to quantum computing, quantum communication, quantum control, and quantum thermodynamics.

\acknowledgements{Special thanks goes to Eric Lutz, who more than a decade ago posed the rather innocent looking question, ``Can we derive the Bremermann-Bekenstein bound by means of Quantum Thermodynamics?" \cite{Deffner2010PRL}. After having developed a more comprehensive framework for \emph{quantum speed limits} the answer can finally be given to be ``yes!". Enlightening discussions with Akram Touil are gratefully acknowledged that helped to streamline the narrative. This research was supported by grant number FQXi-RFP-1808 from the Foundational Questions Institute and Fetzer Franklin Fund, a donor advised fund of Silicon Valley Community Foundation.}


%

\end{document}